\documentclass[pre,twocolumn,showpacs]{revtex4}

\usepackage{amssymb,amsmath}

\usepackage{graphicx}

\begin{document}
\title{Segregation in granular binary mixtures: Thermal diffusion}
\author{Vicente Garz\'{o}\footnote[1]{Electronic address: vicenteg@unex.es}}
\address{Departamento de F\'{\i}sica, Universidad de Extremadura, E-06071
Badajoz, Spain}
\begin{abstract}
A solution of the inelastic Boltzmann equation that applies for
strong dissipation and takes into account non-equipartition of
energy is used to derive an explicit expression for the thermal
diffusion factor. This parameter provides a criterion for
segregation that involves all the parameters of the granular
binary mixture (composition, masses, sizes, and coefficients of
restitution). The present work is consistent with recent
experimental results and extends previous results obtained in the
intruder limit case.
\end{abstract}

\draft
\date{\today}
\maketitle

Segregation and mixing of dissimilar grains is perhaps one of the
most interesting problems in agitated granular mixtures. In some
processes it is a desired and useful effect to separate particles
of different types, while in other situations it is undesired and
can be difficult to control. Several mechanisms for segregation
corresponding to different scenarios have been proposed
\cite{K04}, but the problem is not completely understood yet.
Among the different competing mechanisms, thermal diffusion
becomes one of the most relevant mechanisms at large shaking
amplitude where the sample of macroscopic grains resembles a
granular gas. In this regime, binary collisions prevail and
kinetic theory can be quite useful to analyze the physical
mechanisms involved in segregation processes. Very recently,
Schr\"oter et al. \cite{SUKSS06} have carried out experiments in
agitated mixtures constituted by particles of the same density. To
the best of my knowledge, this is one of the few experiments in
which thermal diffusion has been isolated from the remaining
segregation mechanisms. When convection is practically supressed,
they report results to assess the influence of composition and
dissipation on thermal diffusion. Given that this effect can be
predicted by kinetic theory, in this Letter I study the particle
segregation problem driven by the presence of a thermal gradient
in order to explain some of the trends observed in the above
experiments \cite{SUKSS06}  at large shaking amplitudes. This is
the main motivation of this Letter.

Under the above conditions, the so-called thermal diffusion factor
$\Lambda_{ij}$ characterizes the amount of segregation parallel to
the temperature gradient. Thermal diffusion is caused by the
relative motion of the components of a mixture due to the presence
of a temperature gradient. Due to this motion, concentration
gradients subsequently appear in the mixture producing diffusion
that tends to oppose those gradients. A steady state is finally
achieved in which the separation effect arising from thermal
diffusion is compensated by the diffusion effect. Here, I
determine $\Lambda_{ij}$ from a solution \cite{GD02} of the
inelastic Boltzmann equation that applies for strong dissipation
and takes into account non-equipartition of energy. This latter
feature, not present in the elastic case at equilibrium, has
significant consequences on the transport coefficients of the
granular mixture \cite{GD02}.

The model system considered is a low density binary mixture
composed by smooth inelastic hard spheres ($d=3$) or disks ($d=2$)
of masses $m_i$ and diameters $\sigma_i$. Without loss of
generality, we assume that $\sigma_1>\sigma_2$. The inelasticity
of collisions among all pairs is characterized by three
independent constant coefficients of restitution $\alpha _{11}$,
$\alpha _{22}$, and $\alpha _{12}=\alpha _{21}$. The mixture is in
presence of the gravitational field ${\bf g}=-g \hat{{\bf e}}_z$,
where $g$ is a positive constant and $\hat{{\bf e}}_z$ is the unit
vector in the positive direction of the $z$ axis. In experiments
\cite{SUKSS06}, the energy is usually supplied by vibrating
horizontal walls so that the system reaches a steady state whose
properties are presumed to be, far from the walls, insensitive to
the details of the driving forces. Due to the technical
difficulties involved in incorporating oscillating boundary
conditions, here particles are assumed to be heated by the action
of a stochastic-driving force which mimics a thermal bath.
Although the relation between this driven idealized method with
the use of locally driven wall forces is not completely
understood, it must be remarked that in the case of boundary
conditions corresponding to a sawtooth vibration of one wall the
condition to determine the temperature ratio coincides with the
one derived from the stochastic force \cite{DHGD02}. This suggests
that the results obtained from this driving method are a plausible
first approximation for qualitative comparisons with experimental
results. As will be shown below, the results derived here for the
temperature ratio confirm this expectation.
\begin{figure}
\includegraphics[width=0.7 \columnwidth,angle=0]{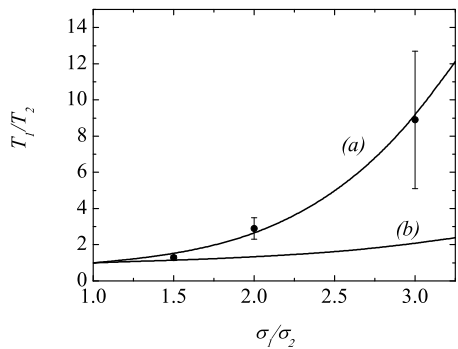}
\caption{Temperature ratio $T_1/T_2$ versus the size ratio
$\sigma_1/\sigma_2$ for $\alpha_{ij}\equiv \alpha=0.78$ in the
case of mixtures constituted by particles of the same mass density
and equal total volumes of large and small particles. The lines
are the kinetic theory results in (a) the stochastic-driving case
and (b) the free cooling case while the points refer to MD
simulations \cite{SUKSS06}. \label{fig0bis}}
\end{figure}

The thermal diffusion factor $\Lambda_{ij}$ ($i\neq j$) is defined
at the steady state in which the mass fluxes ${\bf j}_i$ vanish.
Under these conditions, the factor $\Lambda_{ij}$ is given through
the relation
\begin{equation}
\label{1} -\Lambda_{ij}\nabla \ln T=\frac{1}{x_ix_j}\nabla x_i
,\quad \Lambda_{ij}+\Lambda_{ji}=0,
\end{equation}
where $x_i=n_i/n$ is the mole fraction of species $i$ ($n_i$ is
the number density of species $i$ and $n=n_1+n_2$). The physical
meaning of $\Lambda_{ij}$ can be described by considering a
granular binary mixture held between plates at different
temperatures $T$ and $T'$ under gravity.  For the sake of
concreteness, we will assume that gravity and thermal gradient
point in parallel directions, i.e., the bottom is hotter than the
top. In the steady state (${\bf j}_1={\bf j}_2=0$), Eq.\ (\ref{1})
describes how the thermal field is related to the composition of
the mixture. Assuming that $\Lambda_{12}$ is constant over the
relevant ranges of temperature and composition, integration of
Eq.\ ({\ref{1}) yields $\ln (x_1x_2'/x_2x_1')=\Lambda_{12}\ln
(T'/T)$. Here, $x_i$ refers to the mole fraction of species $i$ at
the plate with temperature $T$ and $x_i'$ refers to the mole
fraction of species $i$ at the plate with temperature $T'$.
Consequently, if $T'>T$ and $\Lambda_{12}>0$, then $x_1'<x_1$
while if $\Lambda_{12}<0$, then $x_1'>x_1$. In summary, when
$\Lambda_{12}>0$, the larger particles accumulate at the top of
the sample (cold plate), while if $\Lambda_{12}<0$, the larger
particles accumulate at the bottom of the sample (hot plate). The
former situation is referred to as the Brazil-nut effect (BNE)
while the latter is called the reverse Brazil-nut effect (RBNE).

\vspace{0.5cm}
\begin{figure}
\includegraphics[width=0.7 \columnwidth,angle=0]{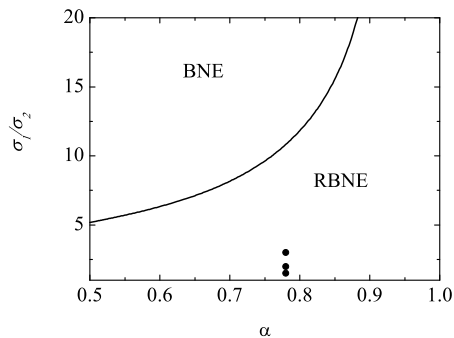}
\caption{Phase diagram for BNE/RBNE for the case studied in Ref.\
\cite{SUKSS06}, namely, mixtures constituted by particles of the
same mass density and equal total volumes of large and small
particles. The data points represent the simulation results for
$\alpha=0.78$ when convection is suppressed. Points below (above)
the curve correspond to RBNE (BNE). \label{fig0}}
\end{figure}

The RBNE was observed first by Hong {\em et al.} \cite{HQL01} in
MD simulations of vertically vibrated systems. They proposed a
very simple segregation criterion that was later confirmed by
Jenkins and Yoon \cite{JY02} by using kinetic theory. More
recently, Breu {\em et al. }\cite{BEKR03} have experimentally
investigated conditions under which the large particles sink to
the bottom and claim that their experiments confirm the theory of
Hong {\em et al.} \cite{HQL01} provided a number of conditions are
chosen carefully.  In addition to the vertically vibrated systems,
some works have also focussed in the last few years on
horizontally driven systems showing some similarities to the BNE
and its reverse form \cite{horizontal}. However, it is important
to note that the criterion given in Ref.\ \cite{HQL01} is based on
some key assumptions: particles are assumed to be elastic,
homogenous temperature and energy equipartition. These conditions
preclude a comparison of the present theory with the above
simulations. Previous theoretical attempts to assess the influence
of non-equipartition on segregation have been recently published.}
Thus, Trujillo {\em et al.} \cite{TAH03} have derived an evolution
equation for the relative velocity of the intruders starting from
the kinetic theory proposed Jenkins and Yoon \cite{JY02} that
applies for weak dissipation. They use constitutive relations for
partial pressures that take into account the breakdown of energy
equipartition between the two species. However, the influence of
temperature gradients which exist in the vibro-fluidized regime is
neglected in Ref.\ \cite{TAH03} because they assume that the
pressure and temperature are constant in the absence of the
intruder. A more refined theory has been recently provided by Brey
{\em et al.} \cite{BMM05} in the case of a single intruder in a
vibrated granular mixture under gravity. The present work covers
some of the aspects not accounted for in the previous theories
\cite{JY02,TAH03,BMM05} since it is based on a kinetic theory
\cite{GD02} that goes beyond the quasi-elastic
limit\cite{JY02,TAH03} and applies for arbitrary composition $x_1$
(and so, it reduces to the results obtained in Ref.\ \cite{BMM05}
when $x_1\to 0$). This allows one to assess the influence of
composition and dissipation on thermal diffusion in bi-disperse
granular gases without any restriction on the parameter space of
the system. This the main value added of this paper and can be
relevant to make comparisons with experiments/simulations in the
dilute regime.

To determine the dependence of the coefficient $\Lambda_{12}$ on
the parameters of the system, we focus our attention on an {\em
inhomogeneous} steady state with zero mass flux and gradients only
along the vertical direction ($z$ axis). Since the flow velocity
vanishes, the momentum balance equation yields $\partial_zp=-\rho
g$, where $p=nT$ is the pressure, $\rho=\rho_1+\rho_2$, and
$\rho_i=m_in_i$ is the mass density of species $i$. To first order
in the gradients, the constitutive equation for the mass flux
$j_{1,z}$ is
\begin{equation}
\label{4} j_{1,z}=-\frac{m_1m_2n}{\rho}D
\partial_zx_1-\frac{\rho}{p}D_p\partial_zp-
\frac{\rho}{T}D'\partial_zT,
\end{equation}
where $D$ is the mutual diffusion coefficient, $D_p$ is the
pressure diffusion coefficient, and $D'$ is the thermal diffusion
coefficient. The condition $j_{1,z}=0$ yields
$\partial_zx_1=(\rho^3/m_1m_2np)(D_p/D)g-(\rho^2/m_1m_2p)(D'/D)\partial_zT$.
Substitution of this relation into Eq.\ (\ref{1}) leads to
\begin{equation}
\label{4.1} \Lambda_{12}=\frac{n\rho^2}{\rho_1\rho_2}\frac{D'-D_p
g^*}{D},
\end{equation}
where $g^*\equiv \rho g/n\partial_zT<0$  is the reduced gravity
acceleration.

Since the mutual diffusion coefficient $D$ is positive
\cite{GD02}, the sign of $\Lambda_{12}$ is determined by the sign
of the quantity $D'-D_pg^*$. Explicit expressions for the
coefficients $D'$ and $D_p$ have been recently obtained from the
Chapman-Enskog method in the first Sonine approximation
\cite{GD02}. Given that the driving-stochastic term does not play
a neutral role in the transport, it must be remarked that the
expressions for the transport coefficients obtained in the driven
case slightly differs from the ones derived in the free cooling
case. In particular, while in the former situation the coefficient
$D'=0$, the thermal diffusion coefficient $D'=-(\zeta/2\nu)D_p$ in
the free case \cite{GD02}. Since the collision frequency $\nu$ and
the total cooling rate $\zeta$ are positive, then
$\text{sgn}(D')=\text{sgn}(-D_p)$. Consequently, the sign of
$\Lambda_{12}$ is the same as that of the pressure diffusion
coefficient $D_p$ in the driving-stochastic case while
$\text{sgn}(\Lambda_{12})=\text{sgn}(-D_p[(\zeta/2\nu)+g^*])$ in
the free case.

Because of $\text{sgn}(\Lambda_{12})=\text{sgn}(D_p)$ in the
driven case, we will focussed henceforth on the explicit
dependence of $D_p$ on the parameters of the system. Its
expression is \cite{GD02}
\begin{equation}
D_{p}=\frac{\rho_{1}p}{\rho^2\nu}\frac{x_2}{x_2+x_1\gamma}
\left(\frac{\gamma}{\mu}-1\right)
,\label{6}
\end{equation}
where $\mu=m_1/m_2$ is the mass ratio, $\gamma=T_1/T_2$ is the
temperature ratio, and $\nu$ is an effective collision frequency
\cite{GD02}.
\begin{figure}
\includegraphics[width=0.7 \columnwidth,angle=0]{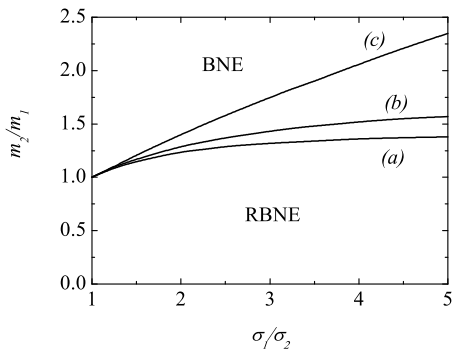}
\caption{Phase diagram for BNE/RBNE in three dimensions for
$\alpha_{ij}=0.7$ and three values of composition: (a) $x_1=0$,
(b) $x_1=0.3$, and (c) $x_1=0.7$. Points below (above) each curve
correspond to RBNE (BNE). \label{fig1}}
\end{figure}

The condition $\Lambda_{12}=0$ (or equivalently, $D_p=0$ in the
driven case) provides the criterion for the transition from BNE to
RBNE. Equation (\ref{6}) shows that the sign of $D_p$ is
determined by the value of the control parameter $\theta\equiv
\gamma/\mu$. This parameter gives the mean-square velocity of the
large particles relative to that of the small particles. Thus,
when $\theta>1$ ($\theta<1$), the thermal diffusion factor is
positive (negative) which leads to BNE (RBNE). The criterion for
the transition condition from BNE to RBNE is $\gamma=\mu$. In the
case of equal granular temperatures (energy equipartition),
$\theta\to \mu^{-1}$ and so, segregation is predicted for
particles that differ in mass, no matter what their diameters may
be. It must be remarked that, due to the lack of energy
equipartition, the condition $\theta=1$ is rather complicated
since it involves all the parameter space of the system. In
particular, even when the species differ only by their respective
coefficients of restitution they also segregate when subject to a
temperature gradient. It is a novel pure effect of inelasticity on
segregation. The criterion for the transition
BNE$\Longleftrightarrow$RBNE is the same as the one found
previously in Ref.\ \cite{TAH03} when $\alpha_{ij}$ is close to 1
and in Ref.\ \cite{BMM05} in the intruder limit case ($x_1 \to
0$). However, as said before, the results obtained here are more
general since they cover all the range of the parameter space of
the system.

To get the explicit dependence of the control parameter $\theta$
on $\alpha_{ij}$, one still needs to compute the temperature ratio
$\gamma$. When the system is driven by means of a stochastic
thermostat, the temperature ratio is determined from the condition
\cite{DHGD02}
\begin{equation}
\label{9} \gamma \zeta_1=\mu \zeta_2 \quad \text{(white-noise
thermostat)},
\end{equation}
where the explicit form of $\zeta_i$ can be found in Ref.\
\cite{GD02}. The condition (\ref{9}) differs from the one derived
in the undriven (free cooling) case \cite{GD99} where $\gamma$ is
obtained by requiring that the partial cooling rates $\zeta_i$
must be equal, i.e.,
\begin{equation}
\label{9.1} \zeta_1=\zeta_2 \quad \text{(free cooling)}.
\end{equation}
This latter condition was used in Ref.\ \cite{BMM05} to determine
the difference of temperatures between an impurity and the
surrounding gas in an open vibrated granular system.

As said above, experiments and molecular dynamics (MD) simulations
have been carried out very recently \cite{SUKSS06} to analyze
segregation in agitated binary granular mixtures composed by
spheres of the same material and therefore, the same mass density
(i.e., $\mu=(\sigma_1/\sigma_2)^3$) and mechanical properties
($\alpha_{ij}=\alpha$). Figure \ref{fig0bis} shows the comparison
of the temperature ratio $T_1/T_2$ between MD simulations
\cite{SUKSS06} and kinetic theory results based on the conditions
(\ref{9}) (driven case) and (\ref{9.1}) (undriven case). The
experimental value of the coefficient of normal restitution is
$\alpha=0.78$ and equal volumes of large and small particles are
taken, i.e., $x_2=(\sigma_1/\sigma_2)^3x_1$. While a good
agreement between kinetic theory and MD simulations is found when
the gas is assumed to be driven by a stochastic thermostat,
significant discrepancies appear in the undriven case, especially
as the size ratio $\sigma_1/\sigma_2$ increases. These results
contrast with the ones obtained in Ref.\ \cite{BMM05} in the
tracer limit ($x_1\rightarrow 0$) where the predictions of
$T_1/T_2$ from kinetic theory based on the condition
$\zeta_1=\zeta_2$ compare quite well with MD simulations. However,
it must noticed that for the cases studied in Ref.\ \cite{BMM05}
the conditions (\ref{9}) and (\ref{9.1}) yield quite similar
results for the dependence of the temperature ratio on the
parameters of the system.

Consider next size segregation driven by thermal difusion. To make
some contact with the recent experimental results of Schr\"oter
{\em et al.} \cite{SUKSS06}, let us consider again
three-dimensional ($d=3$) mixtures constituted by particles with
the same mass density and equal total volumes of large and small
particles. Figure\ \ref{fig0} shows the phase diagram for this
kind of systems. The results show that, for a given value of the
coefficient of restitution, the RBNE is dominant at small diameter
ratios. However, since non-equipartition grows with increasing
diameter ratio, the system shows a crossover to BNE at
sufficiently large diameter ratios. This behavior agrees
qualitatively well with the results reported in Ref.\
\cite{SUKSS06} at large shaking amplitudes where thermal diffusion
becomes the relevant segregation mechanism. At a quantitative
level, we observe that the results are also consistent with the
simulation results reported in \cite{SUKSS06} when periodic
boundary conditions are used to suppress convection since they do
not observe a change back to BNE for diameter ratios up to 3 (see
red squares in Fig.\ 11 of \cite{SUKSS06}). Although the parameter
range explored in MD simulations is smaller than the one analyzed
here, one is tempted to extrapolate the simulation data presented
in Ref.\ \cite{SUKSS06} to roughly predict the transition value of
the diameter ratio at $\alpha=0.78$ (which is the value of the
coefficient of restitution considered in the simulations). Thus,
if one extrapolates from the simulation data at the diameter
ratios of 2 and 3, one sees that the transition from RBNE to BNE
might be around $\sigma_1/\sigma_2=10$, which would quantitatively
agree with the results reported in Fig.\ \ref{fig0}. Figure
\ref{fig0} also shows that the BNE is completely destroyed in the
quasielastic limit ($\alpha \simeq 1$).

Let us now investigate the influence of composition on
segregation. Figure \ref{fig1} shows a typical phase diagram in
the three-dimensional case for $\alpha_{ij}\equiv \alpha=0.7$ and
three different values of the mole fraction $x_1$. The lines
separate the regimes between BNE and RBNE. We observe that the
composition of the mixture has significant effects in reducing the
BNE as the concentration of larger particles increases. In
addition, for a given value of composition, the transition from
BNE to RBNE may occur following two paths: (i) along the constant
mass ratio $m_1/m_2$ with increasing size ratio
$\sigma_1/\sigma_2$, and (ii) along the constant size ratio with
increasing mass ratio $m_2/m_1$. Another aim in this paper is to
assess the influence of dissipation on the phase diagrams. Our
results show that in general the role played by inelasticity is
quite important since the regime of RBNE increases significantly
with dissipation.

In summary, thermal diffusion (which is the relevant segregation
mechanism in agitated granular mixtures at large shaking
amplitudes) has been analyzed in the context of the inelastic
Boltzmann equation by using a kinetic theory that is not
restricted to small dissipation and accounts for the
non-equipartition of energy. The results reported here have been
mainly motivated by recent experimental results \cite{SUKSS06}
focussed on the analysis of three different segregation
mechanisms: void filling, convection, and thermal diffusion.
Concerning the latter effect, our model is able to explain some of
the experimental and/or MD segregation results observed within the
range of parameter space explored. In addition, the theoretical
predictions for the temperature ratio of the mixture obtained from
the driven condition (\ref{9}) agree very well with computer
simulations. A more quantitative comparison with MD simulations in
the low-density regime is needed to show the relevance of the
present theory. As said before, comparison with MD simulations in
the tracer limit case ($x_1\to 0$) \cite{BMM05} has shown the
reliability of the inelastic Boltzmann equation to describe
segregation. Given that the results derived here extends the
description made in Ref.\ \cite{BMM05} to arbitrary values of
composition, one expects that such good agreement is also
maintained for finite values of $x_1$. In this context, it is
hoped that the present results stimulates the performance of such
computer simulations in the dilute regime.

\acknowledgments

 This research has been supported by the
Ministerio de Educaci\'on y Ciencia (Spain) through grant No.
FIS2004-01399, partially financed by FEDER funds.

\end{document}